\documentclass{aa}

\usepackage{xcolor}
\usepackage{graphicx}

\usepackage{txfonts}

\usepackage{hyperref}
\hypersetup{
    colorlinks=true,
    linkcolor=blue,
    filecolor=magenta,      
    urlcolor=blue,
    pdftitle={Overleaf Example},
    pdfpagemode=FullScreen,
    citecolor=blue,
    }

\usepackage{soul}
\usepackage[normalem]{ulem}

\newcommand{\sect}[1]{\text{Sect.~\ref{#1}}}
\newcommand{\figur}[1]{\text{Figure~\ref{#1}}}

\newcommand{\marcs}{\texttt{MARCS}}
\newcommand{\atlas}{\texttt{ATLAS}}
\newcommand{\balder}{\texttt{Balder}}
\newcommand{\multitd}{\texttt{Multi3D}}
\newcommand{\moog}{\texttt{MOOG}}
\newcommand{\iraf}{\texttt{IRAF}}
\newcommand{\qdois}{\texttt{q2}}

\renewcommand{\ion}[2]{\hbox{{#1}{\,\sc #2}}}
\newcommand{\teff}{\mathrm{T}_{\mathrm{eff}}}
\newcommand{\vmic}{\xi_{\mathrm{turb.}}}
\newcommand{\logg}{\log{g}}

\newcommand{\feh}{\mathrm{[Fe/H]}}
\newcommand{\tcond}{\mathrm{T}_{\mathrm{cond}}}

\definecolor{color1}{RGB}{102,153,255}
\definecolor{color3}{RGB}{255,82,210}
\definecolor{color2}{RGB}{0,0,0}
\definecolor{color4}{RGB}{100,179,66}

\begin{document}

   \title{The peculiar composition of the Sun is not related to giant planets}
   
   \author{M. Carlos\inst{1,2}
           \and
           A. M. Amarsi\inst{1}
           \and
           P. E. Nissen\inst{3}
           \and
           G. Canocchi\inst{4}
          }

   \institute{Department of Physics and Astronomy, Uppsala University, Box 516, SE-751 20 Uppsala, Sweden 
            \and
        Observatório Nacional/MCTIC, R. Gen. José Cristino, 77, 20921-400, Rio de Janeiro, Brazil \\
            \email{mariliacarlos@on.br}
            \and
        Department of Physics and Astronomy, Aarhus University, Ny Munkegade 120, DK-8000 Aarhus C, Denmark 
            \and
        Department of Astronomy, Stockholm University, AlbaNova University Center, SE-106 91 Stockholm, Sweden \\ }

   \date{Received Month XX, XXXX; accepted Month XX, XXXX}

  \abstract{Highly-differential spectroscopic studies have revealed that the Sun is deficient in refractory elements relative to solar twins.  To investigate the role of giant planets on this signature, we present a high precision abundance analysis of HARPS spectra for 50 F- and G-type stars spanning $-0.4\lesssim\feh\lesssim+0.5$. There are 29 stars in the sample which host planets of masses $\gtrsim 0.01$ M$_{\mathrm{Jup.}}$.  We derive abundances for 19 elements, and apply corrections to 14 of them for systematic errors associated with one dimensional (1D) model atmospheres, or the assumption of local thermodynamic equilibrium (LTE), or both.  We find that, among the solar twins in our sample, the Sun is Li poor in comparison to other stars at similar age, in agreement to previous studies.
  The sample shows a variety of trends in elemental abundances as a function of condensation temperature.  We find a strong correlation in these trends with $\feh$, with a marginally-significant difference in the gradients for stars with and without giants planets detected, that increases after applying 3D and non-LTE corrections.  
  Our overall results suggests that the peculiar composition of the Sun is primarily related to Galactic chemical evolution rather than the presence of giant planets.
  }


   \maketitle

\section{Introduction}

The first planet discovered outside our Solar System was found around a pulsar by \cite{wolszczan_frail/92}; however, it was the discovery by \cite{mayor_queloz/95} of a planet orbiting a solar-type star that marked the beginning of modern exoplanet research. Since then, more than 5000 planets were detected around different types of stars\footnote{Data from \url{https://exoplanetarchive.ipac.caltech.edu/}}. A detailed review on the first exoplanets observed, the detection techniques and the stellar properties are shown in one of the first reviews on that topic by \cite{udry_santos/07}.

Stellar compositions offer a promising way to investigate and understand how those systems were formed. Stars and their planets form from the same cloud, thus, the elemental abundances of host stars put first-order constraints on the chemical composition of their planet \citep[e.g.][]{thiabaud/15}.   Moreover, planet formation processes can cause small differences in stellar compositions relative to the protostellar nebula \citep[e.g.][]{huhn_bitsch/23}.  Thus, elemental abundances of host stars also have the potential to shed light on how planets form.

Early studies found a positive correlation between stellar iron content $\feh$\footnote{$\feh=\mathrm{A}(\mathrm{Fe})_{\star} - \mathrm{A}(\mathrm{Fe})_{\odot}$ and $\mathrm{A}(\mathrm{Fe})=\log (\mathrm{Fe}) = \log (\mathrm{N}_{\mathrm{Fe}}/\mathrm{N}_{\mathrm{H}}) + 12$, where $\mathrm{N}_{\mathrm{Fe}}$ and $\mathrm{N}_{\mathrm{H}}$ are the number densities of iron and hydrogen respectively.} and the occurrence of massive planets \citep{santos/00, fischer_valenti/05}. The reason for these finds are still on debate, but possible explanations are either that a critical disk accretion rate to form planets correlates with metallicity \citep{liu/16} or  that, instead of a planetary formation cause, we  are observing a dynamical galactic effect due to  stellar migration \citep{haywood/09}.

In this context, the study of other elements might shed more light on planetary formation.
Works from \cite{delgado-mena/10} and \cite{adibekyan/12}, among others, studied a large number of hosting planet stars and enriched the field with important material that was crucial to ignite the discussion about planetary formation.
Following, several studies  tried to link stellar composition and the occurrence of planets 
(\citealt{liu/21}, \citealt{tautvaisiene/22}, and references therein).

Notably, the work of \cite{melendez/09} conclude that the Sun is enriched in volatile elements and
deficient in refractory elements compared to a sample of 11 solar twins (meaning stars with $\Delta\mathrm{T}_{\mathrm{eff},\odot}\lesssim100\,\mathrm{K}$, $\Delta\log{g}_{\odot}\lesssim 0.1\,\mathrm{dex}$, and $\Delta\mathrm{[Fe/H]}_{\odot}\lesssim0.1\,\mathrm{dex}$), none of which to date have confirmed planets. 
A similar result was found in the study of \cite{bedell/18} with a broader sample of solar twins (about 80 stars). One possible explanation is that this material missing from the Sun was used in the formation of terrestrial planets.

In contrast to these results, \cite{gonzalez_hernandez/10} and \cite{gonzalez_hernandez/13} presented a similar study with a sample of about, respectively, 100 and 60 F- and G-type stars (in a broader range of stellar parameters in comparison to solar twins: $5600\leq \teff\, \mathrm{(K)}\leq6400$, $4.0\leq\logg\leq4.6$ and $-0.3\leq\feh\leq+0.5$) and find no evidence correlating the volatile-to-refractory abundance ratio to the presence of rocky planets. More recently, \cite{nibauer/21} analysed a large sample, 1700  solar type stars with $\Delta\mathrm{T}_{\mathrm{eff},\odot}\lesssim195\,\mathrm{K}$, $\Delta\log{g}_{\odot}\lesssim 0.1\,\mathrm{dex}$, and $\Delta\mathrm{[Fe/H]}_{\odot}\lesssim0.1\,\mathrm{dex}$, using APOGEE data and they found two groups of stars regarding their content of refractory elements with the majority of the stars, which includes the Sun, belonging to the refractory poor group.

Several studies have suggested that the peculiar composition of the Sun could be related to planets. \cite{booth_owen/20} argue that the Sun being refractory poor can be explained by the presence of giant planets, that might trap the dust outside their orbits and potentially cause the host star to be refractory poor. 

On the other hand, other works have highlighted complications of this method to find any signatures of planets around stars. \cite{adibekyan/14} and \cite{nissen/15}, which analyse solar type stars, suggest that there is a dependence between the amount of refractory elements and stellar ages. Additionally, \cite{swastik/22} compiled chemical abundances for planet-hosting stars in distinct evolutionary states and conclude that observed abundance trends with planet mass is a consequence of the chemical evolution of our Galaxy. However, \cite{adibekyan/16}, who also found a correlation of refractory elements with stellar ages, suggested that the signature of the chemical evolution of the Galaxy on the amount of refractory elements in hosting planet stars might be erased for stars with similar ages.

Finally, alternative scenarios involving processes in the protoplanetary disk could also explain these abundance patterns, such as a self dust cleansing caused by radiation of the proto-Sun \citep{gustafsson/18b}. Planet engulfments could also cause an excess of  refractory elements in the stellar atmosphere \citep{spina/21,liu/24} as shown in the analysis of wide binary systems presented in \cite{oh/18} and \cite{nagar/20}, for example.

In an attempt to shed light on this topic we present a high precision analysis of 19 elements in high resolution spectra of 50 F- and G-type stars with and without giant planets (\sect{method}). To improve the accuracy of the analysis, we applied corrections to the 1D LTE abundances for 14 elements. We inspect the resulting trends in abundances with condensation temperature (\sect{results}) and find a strong correlation with $\feh$, suggesting that giant planets may at most impart only a second-order effect on the stellar compositions (\sect{discussion}).

\section{Method}
\label{method}

The sample consists of 50 F- and G-type stars observed with the HARPS spectrograph \citep{mayor/03} at the 3.6m ESO telescope. These objects are previously presented in \cite{nissen/14} and \cite{amarsi/19} and can be found in the ESO Science Archive\footnote{\url{http://archive.eso.org/}}. The spectra have high resolution and good signal to noise ($\rm{R}=115000$ and 
$\rm{S/N}\gtrsim 200$). Our study was performed by analysing the red HARPS spectra, which have a wavelength coverage from $5330$ to $6910$ \AA. Additionally, the solar HARPS spectrum was obtained by observing the reflected Sun light in the Vesta asteroid.
Twenty nine stars in our sample host at least one exoplanet ($58$\% of the sample), the exoplanet distances go from $0.03$ to $13.18$ AU with $0.016 \leq \rm{M}\sin i \leq 16.6$\footnote{Data from \url{https://exoplanetarchive.ipac.caltech.edu/}.} $\rm{M}_{\rm{jup.}}$.

Photometric effective temperatures ($\teff$) were calculated by \cite{nissen/14}, while microturbulence velocities ($\vmic$) and iron abundance ($\feh$) are from \cite{amarsi/19}. With the aid of \qdois{} code \citep{ramirez/14}, we updated the surface gravity ($\logg$) values
using the updated magnitudes and parallaxes from GAIA DR3 \citep{gaia_col_a,gaia_col_b} and Yonsei-Yale isochrones \citep{yi/01,kim/02}. We note that the mean difference in $\logg$ from the former values is of about $-0.01$ dex.

The sample consists of stars with $\teff$ from 5450 to 5950 K (with a typical statistical error of 30 K), $\logg$ between 3.9 and 4.6 dex (typical statistical error of $\pm 0.02\,\mathrm{dex}$), $-0.4\leq\feh\leq0.5$ (typical statistical error of $\pm 0.007\,\mathrm{dex}$) and $\vmic$ between $0.7$ and $1.5\,\mathrm{km\,s^{-1}}$ (typical statistical error of $\pm 0.06\,\mathrm{km\,s^{-1}}$). 
The systematic errors are larger, as discussed in \cite{casagrande/10} and \cite{nissen/14}, and can be around $65$ K for $\teff$ and $0.10$ dex for $\logg$.

We calculated the 1D LTE abundances of \ion{Na}{i}, \ion{Mg}{i}, \ion{Al}{i}, \ion{Si}{i}, \ion{S}{i}, \ion{Ca}{i}, \ion{Sc}{i} and \ion{Sc}{ii}, \ion{Ti}{i} and \ion{Ti}{ii}, \ion{V}{i}, \ion{Cr}{i} and \ion{Cr}{ii}, \ion{Mn}{i}, \ion{Co}{i}, \ion{Ni}{i}, \ion{Cu}{i} and \ion{Zn}{i}  measuring the equivalent widths with \iraf{} \citep{doug/86,doug/93} applying differential line by line analysis \citep{bedell/14} using the 2019 version of the 1D radiative transfer code \moog{} \citep{sneden/73}, \qdois{}, and the grid of standard \marcs{} model atmospheres \citep{gustafsson/08}. The line list used for these measurements were adapted from \cite{bedell/18} with hyperfine structure for \ion{S}{i} from \cite{civis/24}, \ion{V}{i}, \ion{Mn}{i}, \ion{Co}{i} from \cite{bedell/18} and \ion{Cu}{i} from \cite{shi/14}. We also determined Li abundances for all stars through spectral synthesis in the same way as in \cite{carlos/19}. We adopt $\teff=5772\,\mathrm{K}$ and $\logg=4.44\,\mathrm{dex}$  for the Sun. For our 1D LTE and 1D non-LTE results, we take $\vmic=1.2\,\mathrm{km\,s}^{-1}$ for the disk-integrated solar flux \citep{takeda/22}.

We considered 3D non-LTE effects for several elements.  First, for Li we adopted the 3D non-LTE versus 1D LTE abundance corrections from \cite{wang/21}.  We note that these corrections are strictly based on $\vmic=1\,\mathrm{km\,s^{-1}}$, and so were careful to adopt that value of $\vmic$ in our 1D LTE analysis of Li.  For C and O, we adopted the 3D non-LTE abundances given in \cite{amarsi/19}; and for Fe we adopted the 3D LTE ones in that same paper, noting that the iron abundances are based on \ion{Fe}{II} lines that show negligible departures from 3D LTE in solar-metallicity FGK-type stars (e.g.~\citealt{amarsi/22}).  For Mg, we adopted the 3D non-LTE abundances given in \cite{matsuno/24}, which are based on 3D non-LTE abundance corrections to the 1D LTE abundances derived in the present study.  For Na, we extended the 3D non-LTE grid presented in \cite{canocchi/24b} and determined absolute 3D non-LTE versus 1D LTE corrections. These were interpolated onto our stellar and solar parameters and subtracted to obtain differential 3D non-LTE versus 1D LTE corrections.  These calculations will be described fully in a forthcoming work (Canocchi et al., in prep.).  All of these 3D non-LTE abundances are based on the code \balder{} \citep{amarsi/18,amarsi/22}, which originates from \multitd{} \citep{leenaarts_carlsson/09}, with several modifications in particular to the equation of state and background opacities.

For Si, S, and Ca, we calculated new 1D non-LTE corrections with the code \balder{}. The calculations were performed on a grid of \marcs{} model atmospheres as described in \citet{amarsi/20}, for five different values of abundance, A(X), and three different values of $\vmic$. This allowed us to calculate a grid of abundance corrections which we then interpolated onto our stellar parameters and 1D LTE abundances using cubic splines.  The exact same approach was adopted for the Sun, and thereby  differential 1D non-LTE abundance corrections were applied on [X/H].
The model atoms for Si originates from \citet{amarsi/17}, with small updates to the hydrogen collisions and complexity as described in \citet{nissen/24}; while the model atom for Ca originates from \citet{asplund/21}, and has been used in \citet{lagae/23}.  For S, we updated the model atom recently employed in \cite{kochukhov/24}, in particular to include electron collision data from OpenADAS \citep{summers/11}.  In any case, our calculations and tests indicate that the differential abundance corrections are negligible at least for the present sample of stars and for the weak subordinate S lines used in this work.  A full description of the model and benchmarking on the Sun will be presented in a forthcoming work (Amarsi et al. in prep.).

In addition, we corrected our abundances of Al, Ti, Cr, Mn and Co for 1D non-LTE departures. For Al, we employed the scripts provided by \cite{nordlander/17}.  For the other elements we used the MPIA non-LTE database \citep{mpia_kovalev/18}, which are based on older model atoms from \cite{bergemann/11}, \cite{bergemann_cescutti/10}, \cite{bergemman_gehren/08} and \cite{bergemann/10} that adopt the Drawin recipe for inelastic collisions with neutral hydrogen \citep{steenbock/84,lambert/93}.  
Lines of \ion{Sc}{i}, \ion{Sc}{ii}, \ion{V}{i}, \ion{Ni}{i}, \ion{Cu}{i}, and \ion{Zn}{i} were treated in 1D LTE.

To evaluate any systematics from the choice of model atmospheres we repeated our 1D LTE analysis with \atlas{} models \citep{castelli_kuruck/04}, for elements from Na to Zn. Ultimately, in a differential analyses the abundance differences between the calculations with \marcs{} or \atlas{} are negligible within the errors. The average difference in abundance is about $-0.005$ dex, and the maximum difference of about $\pm0.01$ dex for the most metal poor stars in our sample ($\feh\leq-0.20$). While \cite{delgado-mena/21} found a significant difference in abundances of C and O when comparing \atlas{} and \marcs{} models, these elements are here adopted from the detailed 3D non-LTE study of \cite{amarsi/19}.

The final abundance errors were estimated by considering the stellar parameters uncertainties and the statistical error associated with the abundances inferred from line to line scatter. For the specific cases of Mg, Cu and Zn, only one line was available. Then, those individual lines were measured three times repeatedly with slightly different continuum positions so as to estimate a statistical uncertainty.  The stellar parameters and planetary information are in  Table \ref{table:1} and  LTE and non-LTE abundances and the respective errors are presented in Table \ref{table:2}, both at the CDS.

\begin{table*}
\caption{Stellar identification, effective temperature (K), microturbulence velocity ($\rm{km\,s}^{-1}$), surface gravity (dex), [Fe/H], age (Gyr) and the respective errors, number of planets detected, and planetary system mass ($\rm{M}_{\rm{jup.}}$) adopted in this work.}              
\label{table:1}      
\centering                             
\begin{tabular}{c c c c c c c c c c c}          
\hline\hline                        

Star & $\teff$ (K) & $\vmic$ ($km\,s^{-1}$) & $\logg$ & $\sigma\,(\logg)$  & $\feh$ (3D LTE) & $\sigma\,(\feh)$ & Age (Gyr)  & $\sigma(\mathrm{Age})$ & Np & PSM ($\mathrm{M}_{\mathrm{Jup.}}$) \\

\hline                                

  HD 1237 &  5507 &  1.25 &  4.520 &  0.016 &   0.2249 &  0.0168 &   2.80 &  1.44 &   1 &   3.370 \\ 
  HD 4208  & 5688 &  1.02 &  4.480 &  0.018 &  $-0.2716$ &  0.0050   & 6.20 &  1.22 &  1  &  0.771 \\
  ...\\

\hline                                             
\end{tabular}
\end{table*}

\begin{table*}
\caption{Stellar identification and stellar abundances in 1D LTE, 1D non-LTE and 3D non-LTE, when available, and their respective errors.}              
\label{table:2}      
\centering                             
\begin{tabular}{c c c c c c c c c c}          
\hline\hline                        

Star & A(Li) 1D LTE & A(Li) 3D non-LTE & $\sigma_+$ & $\sigma_-$ & [C/H] 1D LTE & $\sigma$ &  ... & [Zn/H] 1D LTE & $\sigma$\\

\hline                                

HD 1237 & 2.080 & 2.030 & 0.033 & 0.037 & 0.0900 & 0.0200 & ... & 0.0010 & 0.0120 \\
HD 4208 & $\leq0.48$ & $\leq0.41$ & - & - & $-0.2255$ & 0.0045 & ... & $-0.1900$ & 0.0088 \\
...\\

\hline                                             
\end{tabular}
\end{table*}

\section{Results}
\label{results}

\subsection{Lithium}
\label{lithium}

\begin{figure}
    \centering
    \begin{tabular}{c}
         \includegraphics[width=1\linewidth]{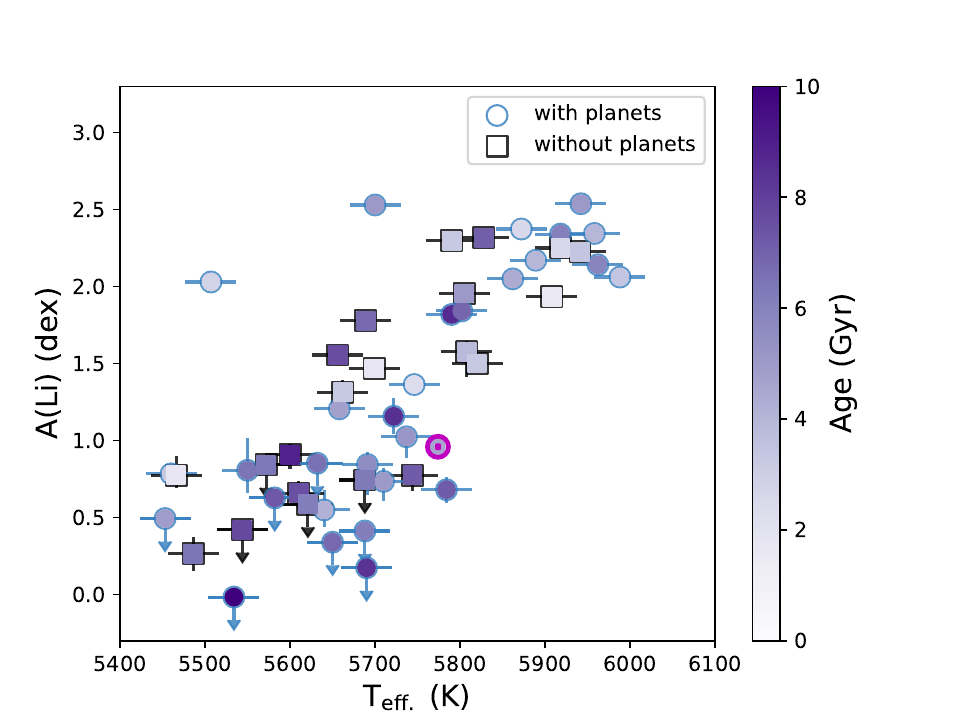} \\ 
    \end{tabular}
    
    \caption{Absolute Li abundance versus $\teff$. Stars with giant planets are shown in blue circles and stars without giant planets detected are presented by black squares, the Sun is shown by its usual symbol in purple. The data is colour-coded by age.}
    \label{fig:ali_teff}
\end{figure}

Because Li burns at low temperatures ($2.5\times10^6\,\mathrm{K}$) below the convective zone of solar type stars, its abundance is highly sensitive to stellar parameters and age \citep{carlos/16}. In addition, in the past two decades works in literature debated a possible connection between Li depletion and the presence of planets. For instance, the work of \cite{israelian/09}, followed by the more refined analysis of \cite{delgado-mena/14}, concludes that the Li depletion is also related to the presence of planets, even if as a secondary effect. Other studies, such as \cite{ghezzi/10} and \cite{llorent_de_andres/24}, also discuss the composition of Li on hosting planet stars. On top of that, the work of \cite{spina/21} claims that a planet engulfment event can cause an excess of Li in the stellar atmosphere.

 With all these considerations, it is hard to assess only the (giant) planet effect on Li abundances in solar type stars with a broader range in stellar parameters, in comparison e.g. solar twins \citep{rathsam/23}. Nonetheless, we present in \figur{fig:ali_teff} our 3D non-LTE Li abundances versus $\teff$ divided into stars with or without giant planets detected, where there is no noticeable distinction between the two sub samples. The typical scatter for a given $\teff$ is of about 0.34 dex, which is the same for 1D LTE abundances and probably caused by other issues (such as age, metallicity, etc.). Moreover, the Sun shows a reasonable Li abundance in comparison to other solar-type stars at given $\teff$ and age; however, if we consider only the solar twins in our sample the Sun is the most Li poor for its age (in agreement with \citealt{carlos/19}). 
 
 From \figur{fig:ali_teff} it is also possible to note two outliers that present higher content of Li for its $\teff$, namely HD 1237 ($\teff=5507\,\mathrm{K}$ and $\mathrm{A(Li)}=2.03\,\mathrm{dex}$) and HD 114613 ($\teff=5700\,\mathrm{K}$ and $\mathrm{A(Li)}=2.53\,\mathrm{dex}$). To assess whether this discrepancy in the Li abundances were a result of their stellar convective envelopes being smaller than expected for its masses and $\feh$ (hence less Li destruction), we calculated the mass of the convective envelope for all stars in our sample by interpolating the values from the YaPSI grid of isochrones \citep{spada/17}. From those values we can speculate that the HD 114613 higher content of Li in comparison to stars at similar $\teff$ might be due to its shallower convective envelope in comparison to other stars at similar $\teff$. On contrary, HD 1237 has a convective envelope mass comparable to other stars at the same $\teff$, thus, their higher content of Li might be due to planet engulfment (as discussed in \citealt{carlos/19}, \citealt{spina/21} and \citealp{sevilla/22}).

\subsection{Stellar abundances and comparison with solar twins}
\label{stellar_abund}

\begin{figure*}
    \centering
    \includegraphics[width=1\linewidth]{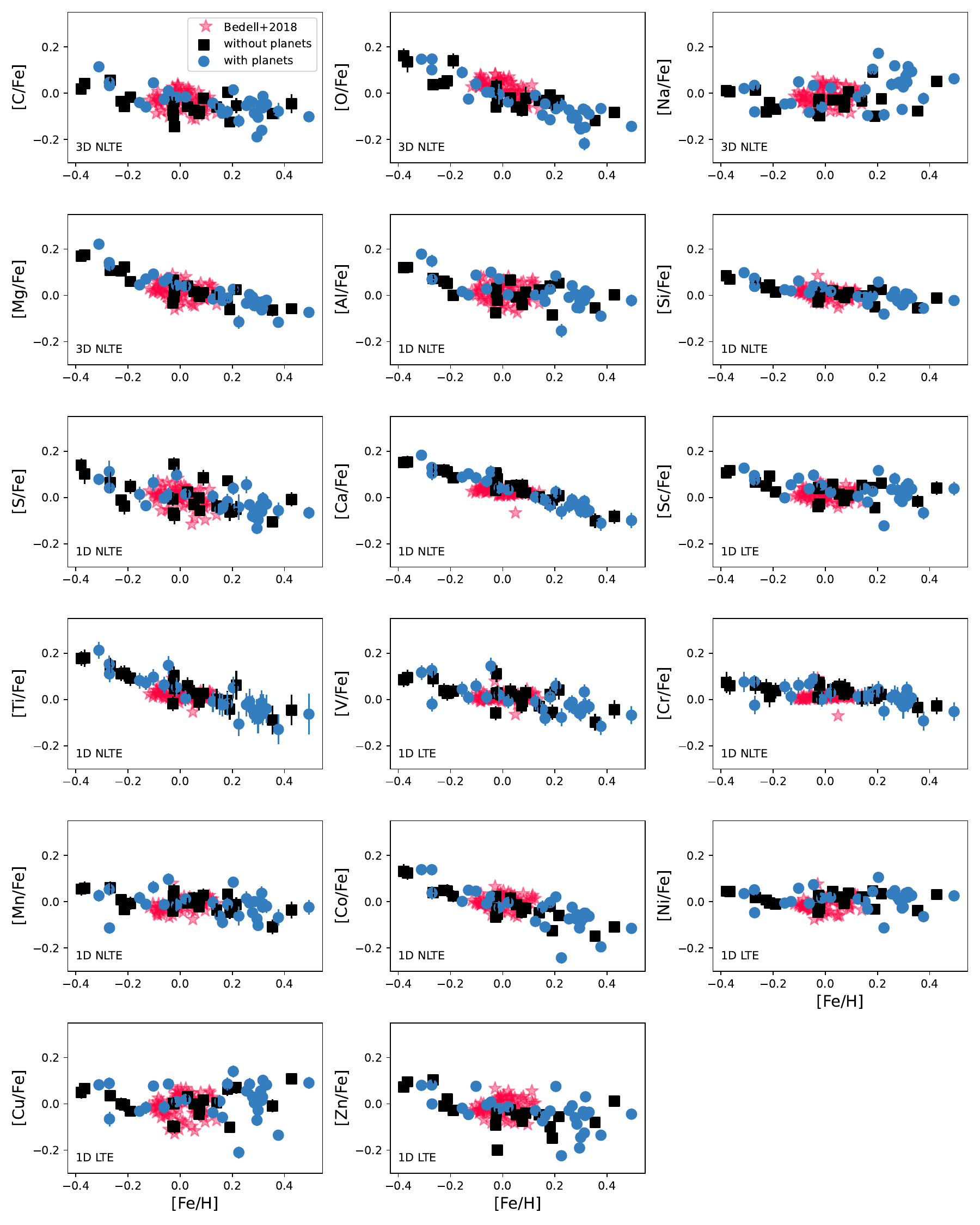}
    \caption{[X/Fe] versus [Fe/H] for stars in our sample with (blue circles) and without (black squares) giant planets detected, in comparison with thin disk solar twins (pink stars) from \protect\cite{bedell/18}.}
    \label{fig:xfe_feh}
\end{figure*}

In this section we discuss  all the other stellar abundances adopted in this work, and what we name the best possible determination: when available we will consider first 3D non-LTE as the best option, followed by 3D LTE, then 1D non-LTE and finally 1D LTE.

For details on how the 3D (non-)LTE analyses of C, O, Mg and Fe affect the data we refer the reader to \cite{amarsi/19} and \cite{matsuno/24}. For the remaining elements, the typical solar-differential corrections vary from $-0.0248$ dex (\ion{Mn}{i}) to $+0.0155$ dex (\ion{Cr}{i}).  For most elements, the 1D or 3D non-LTE corrections tend to diminish the line-by-line scatter, including for \ion{Na}{i} from Canocchi et al. (in prep.), \ion{Al}{i} from \cite{nordlander/17}, and \ion{Si}{i}, \ion{S}{i} and \ion{Ca}{i} calculated in this work.  
However, this was not the case for \ion{Mn}{i} based on corrections from \cite{bergemman_gehren/08}. Similarly, the 1D non-LTE corrections for the \ion{Ti}{i} and \ion{Ti}{ii} from \cite{bergemann/11},  worsened the line-by-line scatter and ionization equilibrium. This may reflect the necessity for full 3D non-LTE analyses, as suggested by \cite{mallinson/24} via their analysis of Ti in metal poor dwarfs in 1D non-LTE and 3D LTE. 
Furthermore, 20\% of the stars in our sample have the line-by-line scatter increased by as much as 0.035 dex for \ion{Cr}{i}, when applying non-LTE corrections from \cite{bergemann_cescutti/10},
strongly affecting their final uncertainties. 
The \ion{Co}{i} line-by-line scatter in 1D LTE was similar to the one after the 1D non-LTE corrections from \cite{bergemann/10}.

\figur{fig:xfe_feh} shows [C/Fe], [O/Fe], [Na/Fe], [Mg/Fe], [Al/Fe], [Si/Fe], [S/Fe], [Ca/Fe], [Sc/Fe], [Ti/Fe], [V/Fe], [Cr/Fe], [Cu/Fe] and [Zn/Fe]  as a function of [Fe/H] for all stars in our sample (with and without giant planets detected) in comparison with the abundances of thin disk solar twins from \cite{bedell/18}; our method of analysis for a given element is highlighted in the bottom left of each panel, with [Fe/H] always in 3D LTE. The Sc, Ti and Cr abundances adopted from now on were calculated as the weighted average from \ion{Sc}{i} and \ion{Sc}{ii}, \ion{Ti}{i} and \ion{Ti}{ii}, and \ion{Cr}{i} and \ion{Cr}{ii} respectively.

The abundances determined in our work are similar to those from \cite{bedell/18}. This good agreement is no surprise since we employed a similar method of analysis and adopted a similar line list. The main difference remains in the fact that \cite{bedell/18} relied predominantly on a 1D LTE analysis.
O is the only exception where they adopted a 1D non-LTE analysis of the \ion{O}{I} 777 nm lines with corrections from \cite{ramirez/13}. Here, our O comes from a 3D non-LTE analysis of the same lines from \cite{amarsi/19}. There appears to be a systematic difference between the two studies such that at solar metallicity the mean [O/Fe] is around 0.04 dex higher in \cite{bedell/18} than in \cite{amarsi/19}.  This means that the Sun is slightly oxygen-poor compared to solar-twins, according to the former study.

It is interesting to note that our [Ca/Fe] decreases with [Fe/H] while the \cite{bedell/18} shows a flat distribution with [Fe/H]. Other works, such as \cite{sun/24} and the GALAH DR4 \citep{buder/24}, also suggest a tendency of decreasing of [Ca/Fe] with [Fe/H].

Moreover, a more careful look at the [Mn/Fe] distributions show that the \cite{bedell/18} values increase with [Fe/H], which is not the case for our data. This somewhat flat distribution in our data only appears after the 1D non-LTE correction (as also found in \citealt{bergemman_gehren/08}), our 1D LTE [Mn/Fe] also increases with [Fe/H]. It is important to note that even after non-LTE corrections the [Mn/Fe] abundances from the GALAH DR3 and GALAH DR4 also rise with [Fe/H] \citep{amarsi/20, buder/24}.

For a detailed comparison between our results and the \cite{bedell/18} ones, we analysed the scatter in the [X/Fe] distributions in the metallicity interval from -0.10 to +0.10 dex. The star-to-star scatter was calculated in bins of 0.05 dex in [Fe/H], and the following values are the weighted mean for  each element. 

Even though our sample has a broader range in $\teff$, $\logg$ and $\vmic$ the star-by-star scatter in our sample is comparable with the scatter from \cite{bedell/18}; and at some cases smaller. Our typical scatter varies from 0.014 dex ([Si/Fe]) to 0.055 dex ([S/Fe]), while the typical scatter in the \cite{bedell/18} data goes from 0.013 dex ([Cr/Fe]) to 0.045 dex ([Cu/Fe]). In general, we present smaller star-by-star scatter for [O/Fe], [Mg/Fe], [Si/Fe], [Co/Fe] and [Cu/Fe]; greater star-by-star scatter for [C/Fe], [Na/Fe], [S/Fe], [Ca/Fe], [Sc/Fe], [Ti/Fe], [V/Fe], [Cr/Fe], [Mn/Fe] and [Zn/Fe]; and similar star-by-star scatter for [Al/Fe] and [Ni/Fe].

\subsection{Condensation temperature trends}
\label{tcs}

\begin{figure*}
\centering
\begin{tabular}{c c}
   \includegraphics[width=0.5\linewidth]{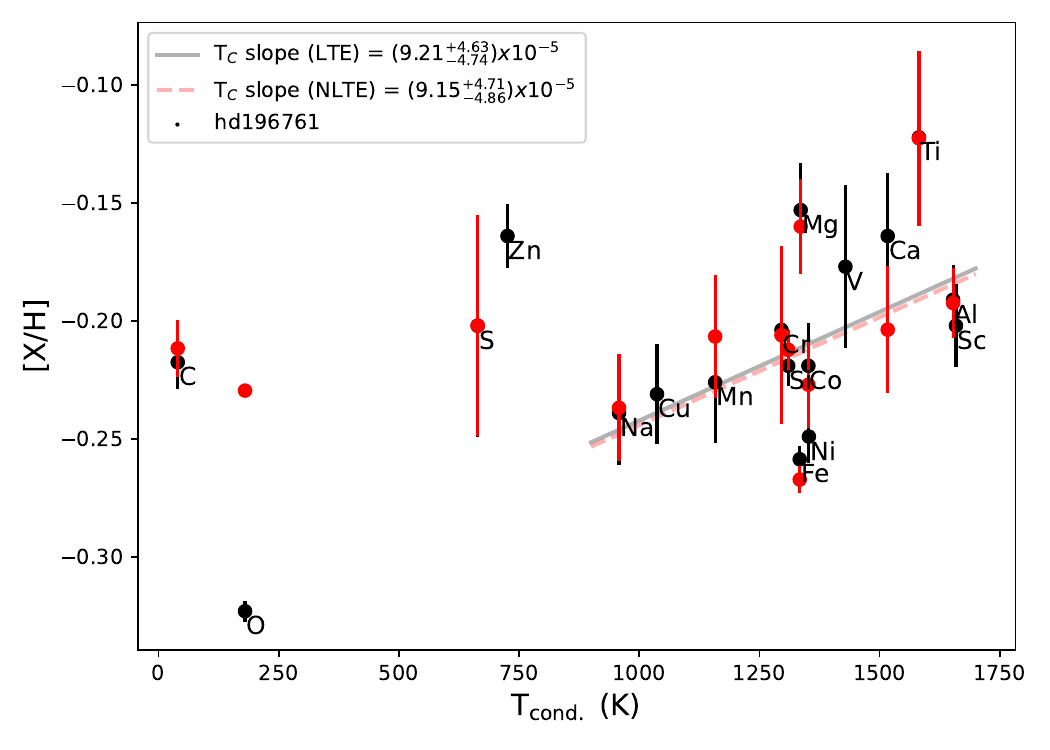}  & \includegraphics[width=0.5\linewidth]{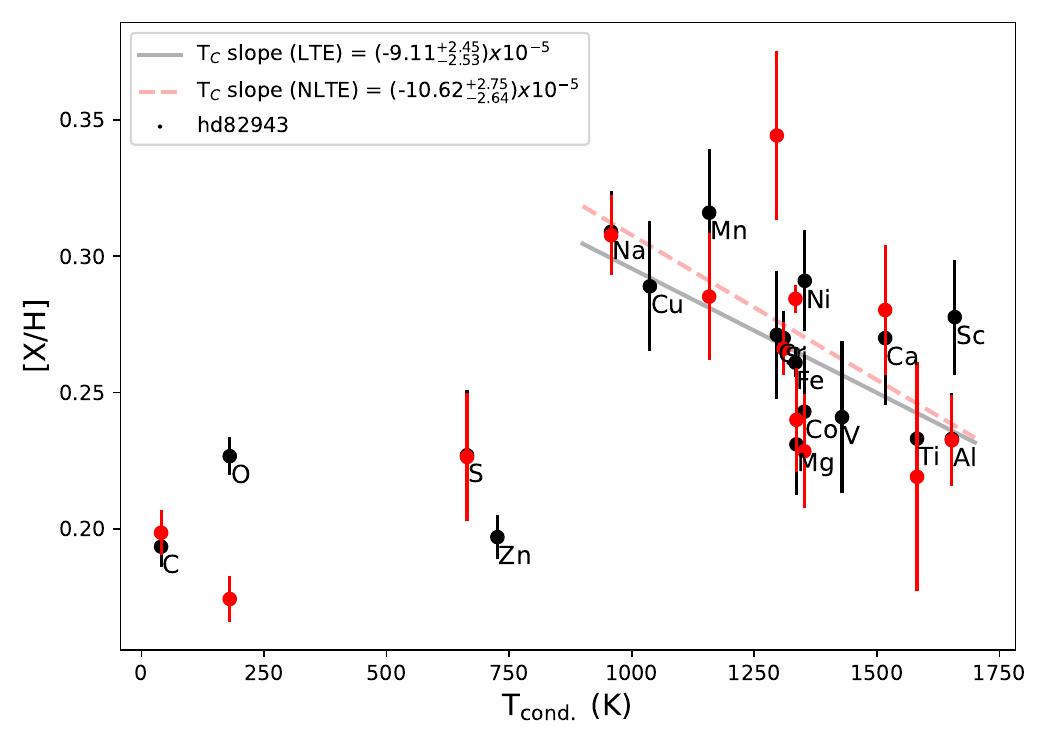}  \\
\end{tabular}

    \caption{[X/H] as a function of elemental condensation temperatures for the stars HD 196761 (without detected giant planets, left panel) and HD 82943 (with two giants planets detected, right panel). Black circles represent elemental abundances in 1D LTE while red circles show 1D non-LTE, 3D LTE or 3D non-LTE (when available). The best fit including only the refractory elements with $\tcond\geq900$ K are represented by the solid grey (1D non-LTE) and red dashed (1D and 3D non-LTE) lines.}
    \label{fig:xh_tcs_example}
\end{figure*}

\begin{figure*}
\centering
\begin{tabular}{c c}
   \includegraphics[width=0.5\linewidth]{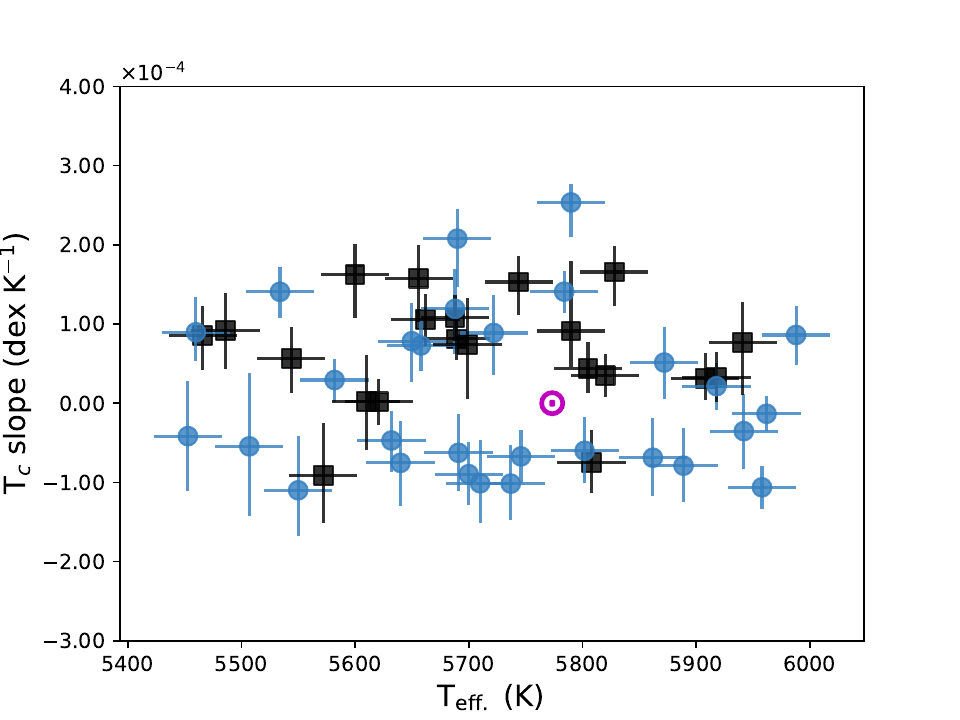}  & \includegraphics[width=0.5\linewidth]{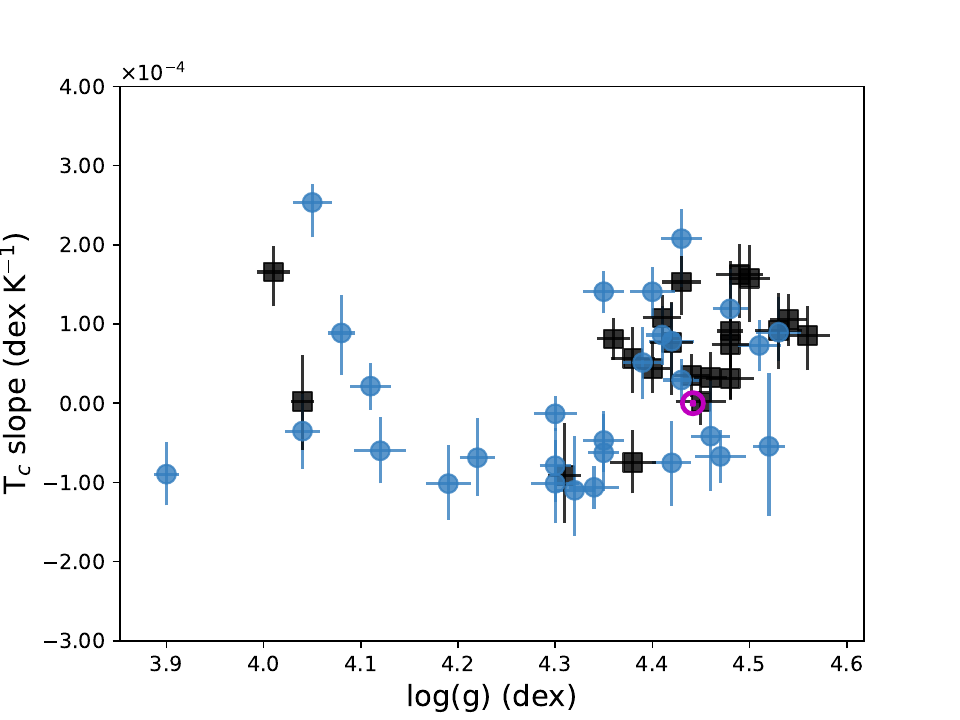}  \\
    \includegraphics[width=0.5\linewidth]{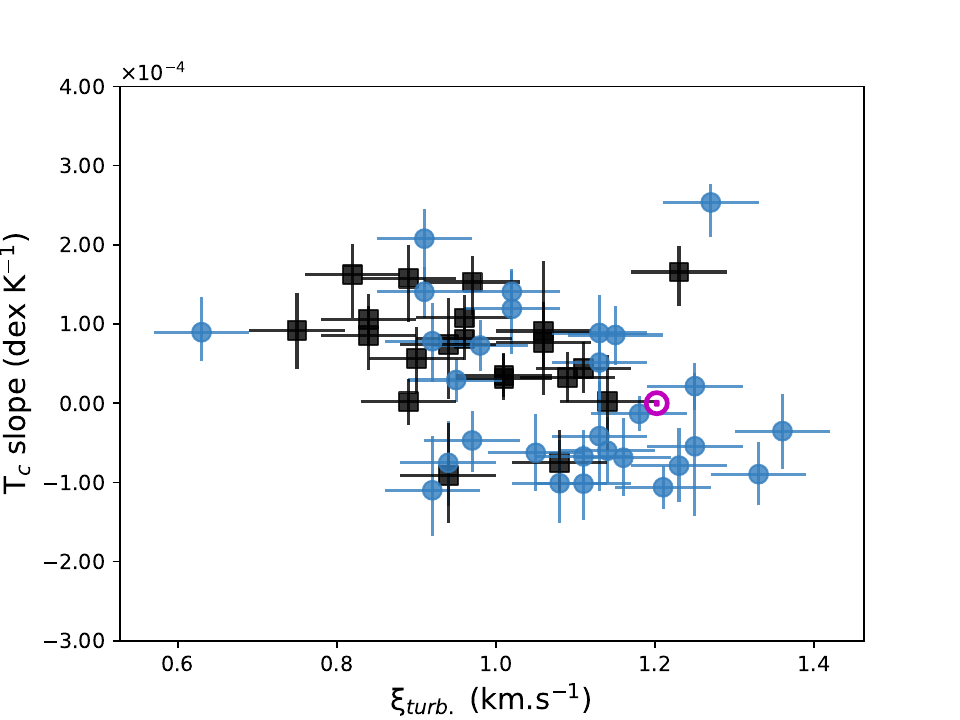} & \includegraphics[width=0.5\linewidth]{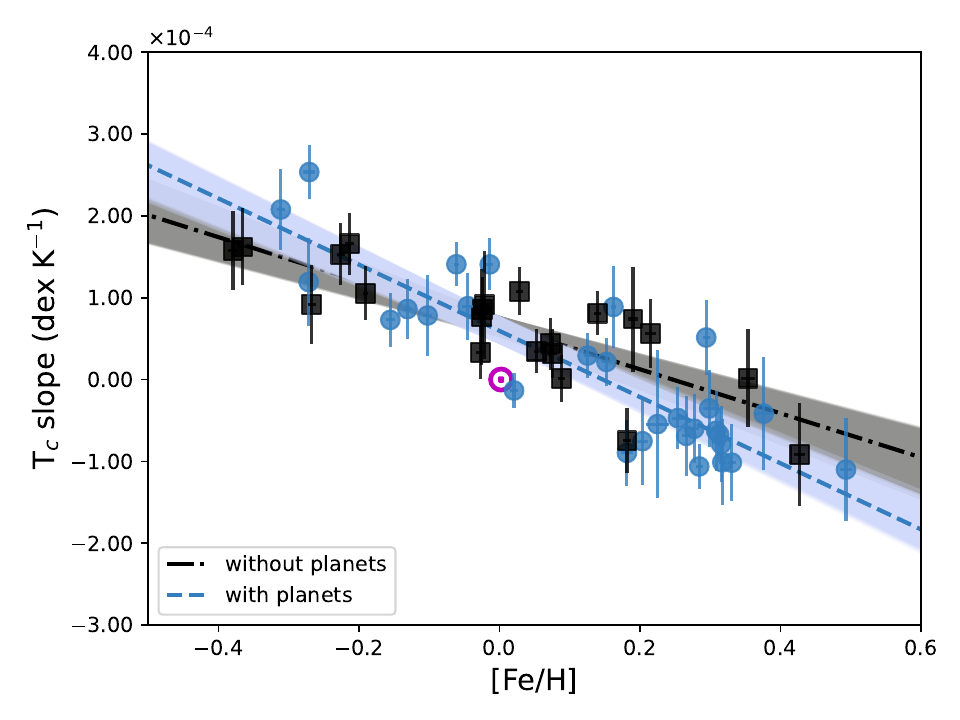} \\ 
\end{tabular}

    \caption{Condensation temperature slopes versus the stellar parameters $\teff$ (top left panel), $\logg$ (top right panel), $\vmic$ (bottom left panel) and $\feh$ (bottom right panel). Stars with giant planets are shown in blue and stars without giant planets detected are presented by black squares, the Sun is shown by its usual symbol in purple.} 
    \label{fig:tcs_sps}
\end{figure*}

\begin{figure}
    \centering
    \begin{tabular}{c}

    \includegraphics[width=1\linewidth]{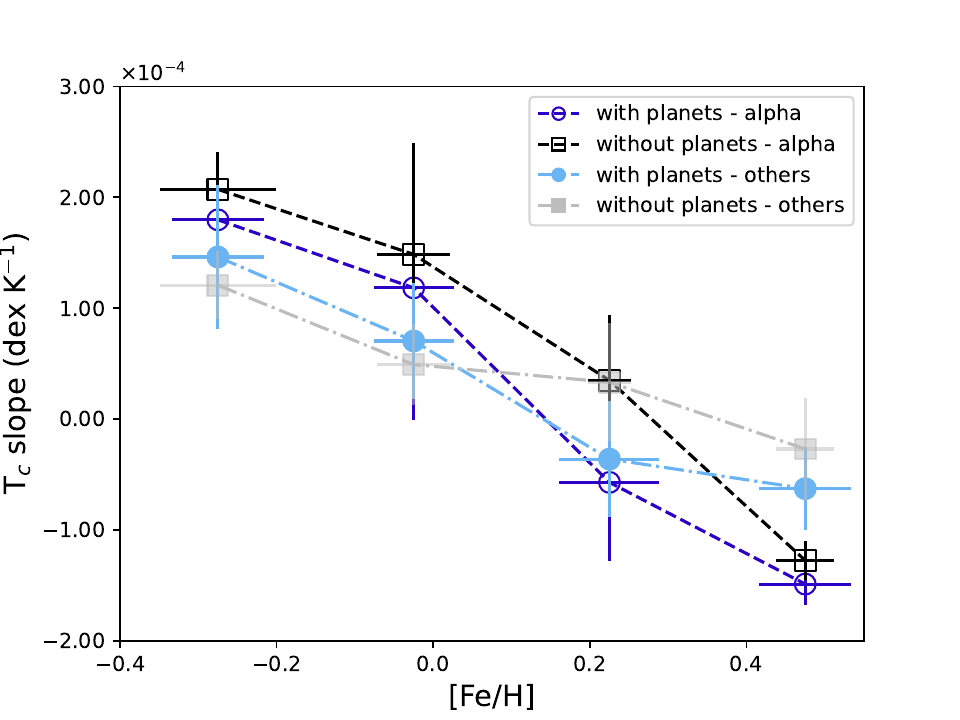} \\
    \end{tabular}
    \caption{$\tcond$ slopes only considering $\alpha-$elements (dashed line) or non $\alpha-$elements (dot-dashed line)  as a function of $\feh$ for stars with (dark and light blue circles) and without (grey and black squares) giant planets.}
    \label{fig:tcs_alpha}
\end{figure}

\begin{figure*}
    \centering
    \begin{tabular}{c c}
    \includegraphics[width=0.5\linewidth]{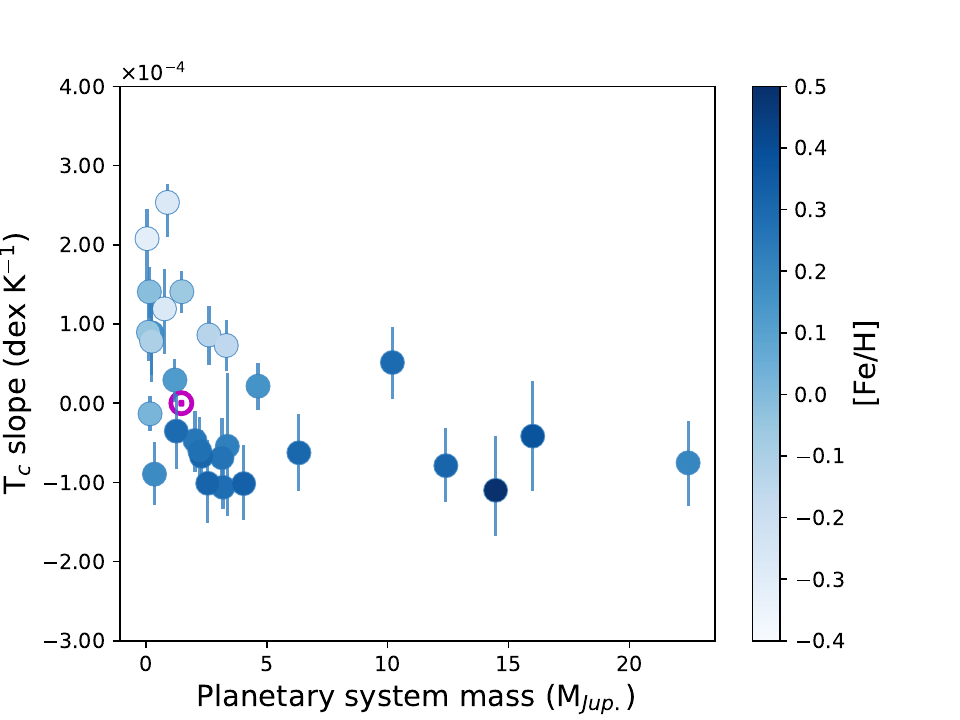}          &
    \includegraphics[width=0.5\linewidth]{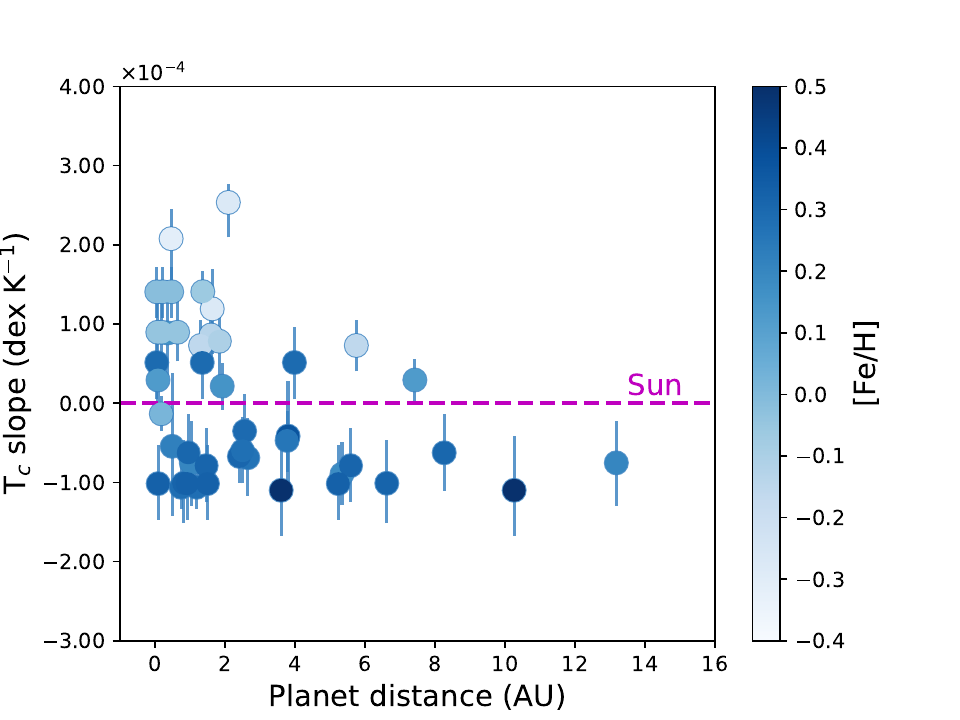} \\
     \includegraphics[width=0.5\linewidth]{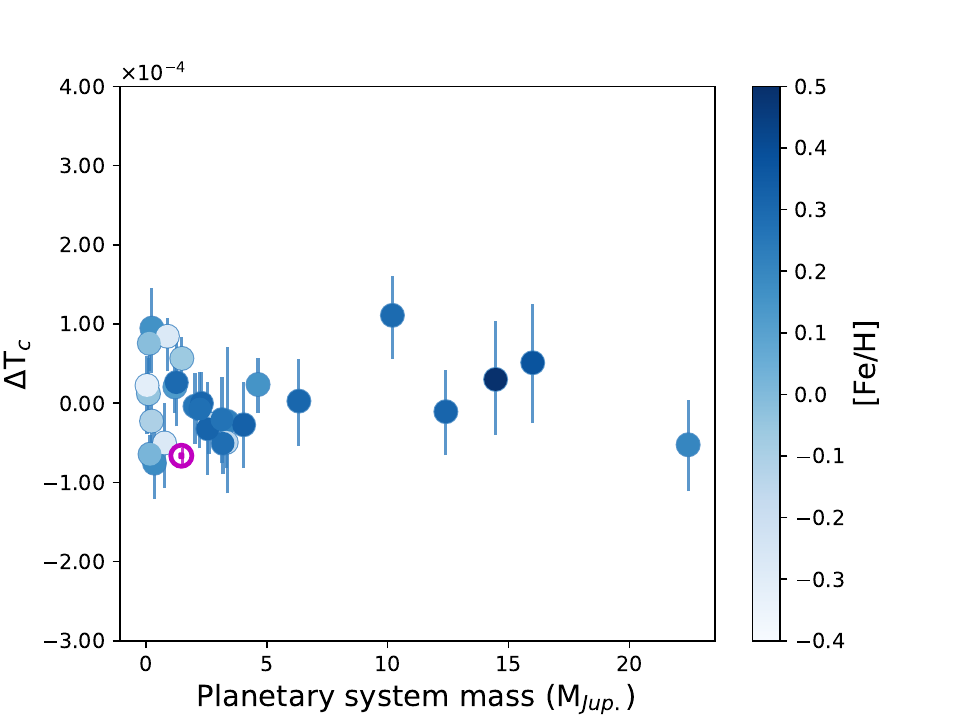}    &
    \includegraphics[width=0.5\linewidth]{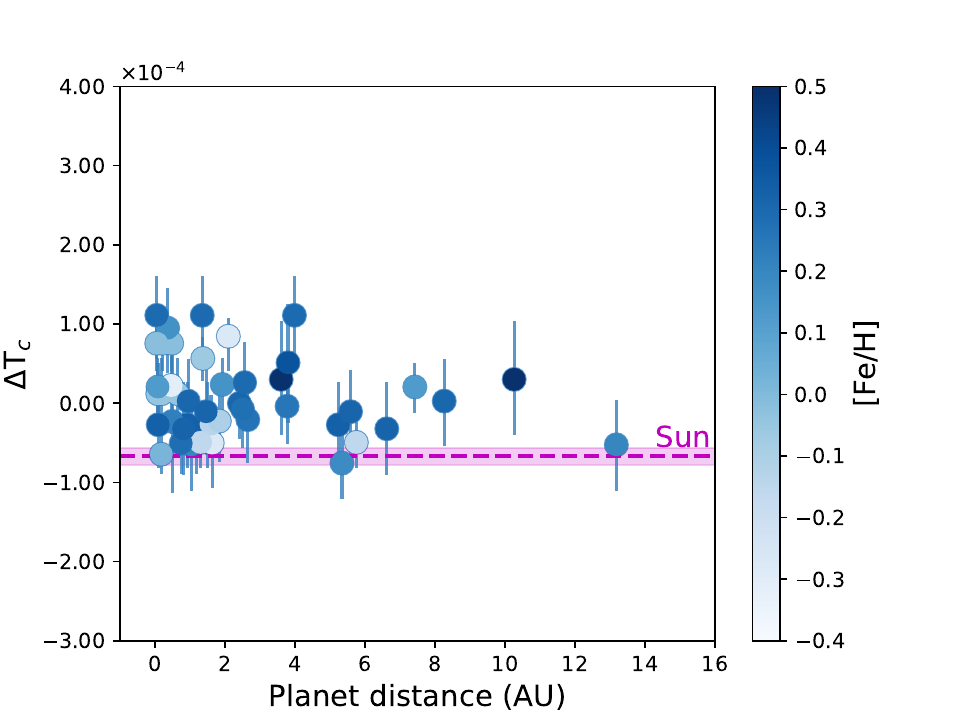} \\ 
    \end{tabular}
    \caption{Top panels: $\tcond$ slopes versus planetary system mass (top left) and planet distance (top right). Lower panels: $\Delta\tcond$ slopes versus planetary system mass (bottom left) and planet distance (bottom right). The Sun is shown by its usual symbol (left panels) or a dashed line (right panels), in purple.}
    \label{fig:tcs_massp}
\end{figure*}

\begin{table*}
\caption{Stellar identification, $\tcond$ slopes (dex K$^{-1}$) using the ``best'' set of abundances and the 1D LTE ones, and the respective errors. }              
\label{table:3}      
\centering                             
\begin{tabular}{c c c c c c c}          
\hline\hline                        

Star & $\tcond$ slope (best) & $\sigma+$ & $\sigma-$ & $\tcond$ slope (1D LTE) & $\sigma+$ & $\sigma-$ \\

\hline                                

  HD 1237 &  $-5.4459\times10^{-5}$ &  $9.4150\times10^{-5}$ &  $8.6695\times10^{-5}$ & $-1.4942\times10^{-5}$ &    $7.6505\times10^{-5}$  &  $7.9058\times10^{-5}$ \\
  HD 4208 &  $1.1913\times10^{-4}$ &  $4.9421\times10^{-5}$  &  $5.6359\times10^{-5}$  &  $1.4462\times10^{-4}$ &   $4.2878\times10^{-5}$  &  $5.5120\times10^{-5}$ \\
...\\

\hline                                             
\end{tabular}
\end{table*}

\figur{fig:xh_tcs_example} presents [X/H] as a function of elemental condensation temperatures ($\tcond$, from \citealt{lodders/03}) for two stars in our sample. For comparison we plot both the 1D LTE (black) and non-LTE (red), either 1D or 3D, abundances. And, in order to assess the behaviour of those abundances distributions for each star, we decided to perform two linear fits: one considering all abundances in 1D LTE, and the other considering the ``best'' available abundances for a given element, namely 3D non-LTE, 3D LTE, 1D non-LTE, or 1D LTE as applicable. The best linear fit for elements with $\tcond\geq 900 \, \mathrm{K}$ was retrieved considering the maximum likelihood estimation. The choice to consider only elements with $\tcond\geq 900 \, \mathrm{K}$ for the linear fits is discussed in detail in \cite{bedell/18}. The uncertainties were estimated through Markov Chain Monte Carlo (MCMC) using the python package \texttt{emcee} \citep{emcee}. 

The $\tcond$ slopes derived in this way for each case and their respective errors are present in Table \ref{table:3} at the CDS. We note that the two linear fit cases (1D LTE and ``best'' abundances) are compatible in one sigma for every star.
From now on the discussion about the $\tcond$ slopes are restricted to those considering the ``best'' set of abundances.

In \figur{fig:tcs_sps} we present the $\tcond$ slopes for stars with and without giant planets, versus stellar parameters $\teff$ (top left), $\logg$ (top right), $\vmic$ (bottom left) and $\feh$ (bottom right). No evident correlation with $\teff$, $\logg$ and $\vmic$ was found.  On the other hand, a strong anti-correlation with $\feh$ is seen on the bottom right panel of \figur{fig:tcs_sps}. Once more, the best linear fits, for hosting giant planets stars (blue dashed line) and stars without detected giant planets (black dot-dashed line), were retrieved considering the maximum likelihood estimation with the uncertainties  computed through MCMC.  This time we took into account the $\tcond$ slopes and their errors, and $\feh$. The blue and black dashed area represents the $1.5\sigma$ interval for each linear fit.

Moreover, from the bottom right panel of \figur{fig:tcs_sps}, it is noticeable that the Sun is marginally refractory poorer than stars at similar metallicities, being $1.7$ standard deviation below the stars distributed within $-0.05\leq\feh\leq+0.05$. To test if the fact that the Sun seems refractory poor is due to systematic errors in the stellar parameters we perturbed $\teff$, $\logg$ and $\vmic$ until the stars in our sample were shifted down so the Sun was the middle of the distribution. No significant changes in this distribution were found for $\teff$ and $\logg$ when using perturbations of up to $300\, \mathrm{K}$ and $0.3\, \mathrm{dex}$. For $\vmic$ a difference of $1\,  \mathrm{km\,s}^{-1}$ was needed so the Sun can be considered typical in this sample.

The $\tcond$ slopes versus $\feh$ trend presumably reflects Galactic chemical evolution.  To gain a deeper understanding of it, we repeated the steps above but separating elements into their two major production sites \citep{nomoto/13}: first, considering only alpha elements (O, Mg, Si, S, Ca and Ti); and secondly, considering non-alpha elements. \figur{fig:tcs_alpha} presents the comparison between the different $\tcond$ slopes as a function of $\feh$, both for stars with or without giant planets detected. The alpha and non-alpha slopes differ typically in $0.41\times10^{-4} \, \mathrm{dex\,K}^{-1}$ and $0.93\times10^{-4} \, \mathrm{dex\,K}^{-1}$, for stars with and without giant planets detected respectively. The slopes for stars with or without giant planets detected differ typically in $0.29\times10^{-4} \, \mathrm{dex\,K}^{-1}$ and $-0.07\times10^{-4} \, \mathrm{dex\,K}^{-1}$, when we consider only alpha elements or when we consider non-alpha elements respectively.
Although a variation was found, the difference in the $\tcond$ slopes for alpha or non-alpha elements, and also for stars with or without giant planets detected, are compatible in $1\sigma$ for most $\feh$ values.

The chemical evolution of our Galaxy may also impart a signature on the behaviour of $\tcond$ slopes versus planetary system mass and planet distance.  These are shown in the top left and right panels of \figur{fig:tcs_massp} respectively.  A weak anti-correlation was found for  the $\tcond$ slope versus planetary system mass and planet distances (Spearman coefficient equal to $-0.60$ and $-0.58$, respectively), where the stellar systems with more massive and distant planets present $\tcond$ slopes lower than $1.0\times10^{-4}\, \mathrm{dex\,K}^{-1}$.  This is in agreement with \cite{yun/24}, who analysed thin disk stars APOGEE DR17 spectra and also found that for giant planets the $\tcond$ slopes were negative for most of the stars.
Interestingly, this anti-correlation goes away in our study in the lower panels of \figur{fig:tcs_massp}, which replace the $\tcond$ slope with the $\Delta\tcond$ slope, given by the $\mathrm{T}_{\mathrm{cond,obs}}$ slope $-$ the $\mathrm{T}_{\mathrm{cond,fit}}$ slope, in other words subtracting the 
linear fit as a function of $\feh$ (blue dashed line in \figur{fig:tcs_sps}).
We also notice that in these plots the Sun is refractory poor when compared to other solar like stars.

\section{Discussion and conclusions}
\label{discussion}

There is a strong correlation between the $\tcond$ slopes and $\feh$ shown in \figur{fig:tcs_sps}. This suggests that studies of abundance trends with condensation temperature are strongly affected by the chemical evolution of our Galaxy.
The work of \cite{adibekyan/14} analysed solar analogues and found a dependence between $\tcond$ slopes and stellar age and suggested that the $\tcond$ slopes are affected by the chemical evolution of our Galaxy, although they were not able to find any relation between the $\tcond$ slope and $\feh$. More recently, \cite{sun/24} found a correlation between the $\tcond$ slope and $\feh$ for stars with $-0.4\lesssim \feh \lesssim+0.4$, but no significant correlation were found for solar twins.  We also found that the corrections to the 1D LTE abundances only lightly affect the  $\tcond$ slopes, as discussed in \sect{tcs}.  Thus, the $\tcond$ slope versus $\feh$ plot with only 1D LTE abundances is very similar to the one considering non-LTE corrections, shown in the bottom right panel of \figur{fig:tcs_sps}. This shows that the correlation the $\tcond$ slopes and $\feh$ seen is this figure is not related to departures from 1D LTE.

Giant planets may impart a second order effect on abundance trends with condensation temperature, given the marginally-significant differences in the linear correlations of the $\tcond$ slope versus $\feh$ for the two populations of stars (with and without giant planets), overplotted in \figur{fig:tcs_sps}. Increasing the sample of high resolution spectra of stars with and without planets would help us make stronger statement regarding the differences between the two linear fits in this panel. Moreover, improving the accuracy of abundance analyses appears to be important.  The difference between the slopes for stars with and without giant planets is slightly affected by the corrections to 1D LTE abundances. The difference is increased from $1\sigma$ in 1D LTE, to $1.5\sigma$ when 3D/non-LTE corrections are applied.

Finally, the Sun is slightly refractory poor when compared to other stars of similar $\feh$ ($-0.05\lesssim \feh \lesssim+0.05$), with marginal significance. This is in agreement with analyses of solar twins \citep{melendez/09,bedell/18} and solar analogues \citep{rampalli/24}. Nine stars are within this $\feh$ interval (including the Sun), of which four of them have giant planets detected.  In this region there is no obvious difference in the correlation of the $\tcond$ slope versus $\feh$ (\figur{fig:tcs_sps}) for stars with and without giant planets. This suggests that the Sun is peculiar mainly for some other reason than the presence and influence of giant planets.

\section*{Data availability}

Tables 1, 2 and 3 are available at the CDS.

\begin{acknowledgements}
MC and GC acknowledges funds from the Knut and Alice Wallenberg foundation. This work is a part of the project “Probing charge- and mass-transfer reactions on the atomic level” supported by the Knut and Alice Wallenberg Foundation (2018.0028). 
AMA acknowledges support from the Swedish Research Council (VR 2020-03940) and from  the Crafoord Foundation via the
Royal Swedish Academy of Sciences (CR 2024-0015).  
This research was supported by computational resources provided by the Australian Government through the National Computational Infrastructure (NCI) under the National Computational Merit Allocation Scheme and the ANU Merit Allocation Scheme (project y89).
This work presents results from the European Space Agency (ESA) space mission Gaia. Gaia data are being processed by the Gaia Data Processing and Analysis Consortium (DPAC). Funding for the DPAC is provided by national institutions, in particular the institutions participating in the Gaia MultiLateral Agreement (MLA). The Gaia mission website is \url{https://www.cosmos.esa.int/gaia}. The Gaia archive website is \url{https://archives.esac.esa.int/gaia}. This research has made use of the NASA Exoplanet Archive, which is operated by the California Institute of Technology, under contract with the National Aeronautics and Space Administration under the Exoplanet Exploration Program.
\end{acknowledgements}

\bibliography{references}
\bibliographystyle{aa}

\end{document}